# Enumeration of Hamiltonian Cycles in 6-cube

September 25, 2018

Michel Deza[1] and Roman Shklyar[2]

**Abstract**

Finding the number $2H_6$ of directed Hamiltonian cycles in 6-cube is problem 43 in Section 7.2.1.1 of Knuth's *The Art of Computer Programming* ([Kn10]); various proposed estimates are surveyed below. We computed exact value:

$H_6 = 14,754,666,508,334,433,250,560 = 6!*2^4*217,199*1,085,989*5,429,923$.
Also the number $Aut_6$ of those cycles up to automorphisms of 6-cube was computed as 147,365,405,634,413,085

A *Hamiltonian cycle* in a graph is a cycle that visits each vertex exactly once. Let $H_n$ denote the number of Hamiltonian cycles in *n-cube*(the graph of n-dimensional hypercube). An *automorphism* of a graph is a permutation of its vertex-set preserving its edge-set. Let $Aut_n$ denote the number of Hamiltonian cycles in n-cube up to the group of automorphisms of n-cube. Let $Weight_n$ denote the number of Hamiltonian cycles in n-cube up to the *weight*. (This another equivalence is well explaned in (citePa01).

The number $H_n$ is given for $n \leq 5$ in OEIS (On-Line Encyclopedia of Integer Sequences) as the sequence A066037: number of Hamiltonian cycles in the binary n-cube, or the number of cyclic n-bit Gray codes ([Sl08]). In fact, $H_n = \frac{1}{2}OH_n$, where $OH_n$ is given for $n \leq 5$ by the sequence OEIS A003042: number of *directed* Hamiltonian cycles (or Gray codes) on n-cube. Finding $OH_6$ is problem 43 in Section 7.2.1.1 of Knuth's *The Art of Computer Programming* ([Kn10]). In Volume 4, exercises, there Knuth also improved the general lower bound to $H_n \geq (\frac{n}{4e} - O(log^2 n))^{2^n}$.

---

[1]Michel.Deza@ens.fr, École Normale Supérieure, Paris, and JAIST, Ishikawa
[2]romans@ariel.ac.il, Ariel University Center of Samaria

Also, $H_n = \frac{n!}{2}EH_n$, where $EH_n$ is given for $n \leq 5$ by the sequence OEIS A091302: number of equivalence classes of Hamiltonian cycles (or Gray codes) in the binary n-cube. In fact, $H_n$ is a multiple of $\frac{n!}{2}$ since, representing the vertices of n-cube as binary $n$-sequences, any directed cycle starting from the sequence of $n$ zeroes induces a permutation on the $n$ bits, namely the order in which they first get set to 1.

The problem to find $H_n$ was originated by Gilbert in 1958 ([Gi58]). Perezhogin and Potapov, 2001 ([PePo01]) proved that $a_n \leq H_n \leq b_n$ for $n \to \infty$, where $a_n = e^{2^{n-1}(ln(n)-1+o(1))}, b_n = e^{2^n(ln(n)-1+o(1))}$. Let $M_n$ denote the number of perfect matchings of n-cube; its values up to $n = 6$ are given in OEIS by the sequence A005271. Above bound implies that $\lim_{n\to\infty} \frac{log(H_n)}{log(M_n)} \in [1,2]$. Feder and Subi, 2009 ([FeSu09]) proved that this limit is 2; so,it holds $H_n = M_n^{2-o(1)}$.

For $n \leq 6$ the present state of art is given in the Table below.

For $n = 2$ and 3 this Table can be easily filled by hand. The values $H_n$, $Aut_n$ and $Weight_n$ were obtained by Parhomenko in 2001 ([Pa01]) for $n = 4$ and by Dejter and Delgado in 2007 ([DeDe07]) for $n = 5$. The value $Weight_6$ was obtained by Chebiryak and Kroening in 2008 ([ChKr08]). The exact value $H_6 = 14,754,666,508,334,433,250,560$, computed by us, as well as $M_5^2$ and $M_6^2$, are given in the Table by upper round.

| n | $H_n$ | $M_n^2$ | $Aut_n$ | $Weght_n$ |
|---|---|---|---|---|
| 2 | 1 | 4 | 1 | 1 |
| 3 | 6 | 81 | 1 | 1 |
| 4 | 1,344 | 73,984 | 9 | 4 |
| 5 | 906,545,760 | $3.471 * 10^{11}$ | 237,675 | 28 |
| 6 | $1.475 * 10^{22}$ | $2.667 * 10^{26}$ | $1,473 * 10^{17}$ | 550 |

In order to develop the computation solution, we used the BGL(Boost Graph libraries ) the C++ library, developed by Siek, Lie-Quan Lee and Lumsdaine ([SQL01]) which involves easy graph construction, implementation and the effective parallel computing. The program was written in the Microsoft Visual Studio 2010 developing environment, which includes new and effective solutions for developers building applications (both managed and native) that take advantage of multiple cores. By the right memory handling and process parallelization the computing time lasts about of 6 months.

Note that $H_3$=3!, $H_4 = 4! * (2^3) * 7$, $H_5 = 5! * (2^2) * 617 * 3,061$ and $H_6$=6!*$2^4$*217,199*1,085,989*5,429,923; so, the integer $\frac{H_n}{n!}$ with n=3,4,5,6 has exactly n-3 odd prime divisors.

In the next Table we list known upper bounds of $H_6$. Two last bounds are upper rounds obtained from asymptotic bounds ([?] and [FeSu09]) by replacing o(1) by 1 and 0, respectively. The absolute error in the upper bound, obtained by Feder and Subi, is going to zero when $n \to \infty$ and it became very small even for the small values of n. So, the absolute upper bound as about equal to the exact value of $H_n$ and the computing the $H_n$ for $n \leq 7$ ,perhaps, is not of big practical importance.

| Known upper bounds of $H_6$ | upper round |
|---|---|
| Dixon and Goodman, 1975 [DiGo75] | $1.5 * 10^{40}$ |
| Douglas, 1977 [Do77] | $1.1 * 10^{35}$ |
| Silvermann et al., 1983 [SVS83] | $3.7 * 10^{29}$ |
| Clark, 2000 [Cl00] | $1.3 * 10^{30}$ |
| Perezhojin and Potapov, 2001 [PePo01] | $4.1 * 10^{24}$ |
| Feder and Subi, 2009 [FeSu09] | $2.7 * 10^{26}$ |
| **obtained value of $H_6$** | $1.4 * 10^{22}$ |

The edge-set of 2n-cube can be partitioned into $n$ Hamiltonian cycles ([ABS90]). To each such partition it correspond $2^{n-1}$ *Hamilton orientations* of 2n-cube obtained by orienting one of the cycles and selecting one of 2 possible orientations on each of remaining $n-1$ cycles. In forthcoming paper we enumerate Hamilton orientations of 4- and 6-cube and study quasimetrics associated with each orientation.

# References

[ABS90] B.Alspach, J.C.Bermond and D.Sotteau, *Decompositions into cycles I: Hamilton decompositions*, in *Cycles and rays*, G.Hahn et al. (eds.) Kluwer Academic Press (1990) 9–18.

[ChKr08] Y.Chebiryak and D.Kroening, *Towards a Classification of Hamiltonian Cycles in the 6-Cube*, J.Satisfiability, Boolean Mod. and Comp., **4** (2008) 57–74.

[Cl00] L.H. Clark, *A new upper bound for the number of Hamiltonian cycles in the n-cube*, J. Comb. Infor. System Sci., **25** (2000) 3–37.

[DeDe07] I.Dejter and A.A.Delgado, *Classes of Hamilton Cycles in the 5-cube*, J. Comb. Math. Comput., **61** (2007) 81–95.

[DiGo75] E.Dixon and S.Goodman, *On the number of Hamiltonian circuits in the n-cube*, Proc. Amer. Math. Soc., **50** (1975) 500-504.

[Do77] R.J. Douglas, *Bounds on the number of Hamiltonian cirucits in the n-cube*, Discrete Math., **17** (1977) 143-146.

[FeSu09] T.Feder and C.Subi, *Nearly Tight Bounds on the Number of Hamiltonian Circuits of the Hypercube and Generalizations*, Information Processing Letters, **109** (2009) 267–272.

[Gi58] E.N. Gilbert, *Gray codes and paths on the n-cube*, Bell System Tech. J., **37-3** (1958) 815–826.

[Kn10] D.E. Knuth, *The Art of Computer Programming*, vol. 4A, Combinatorial Algorithms (to appear).

[Pa01] P.P. Parkhomenko, *Classification of the Hamiltonian Cycles in Binary Hypercubes*, Automation and Remote Control, **62-6** (2001) 978–991.

[PePo01] A.L. Perezhogin and V.N. Potapov, *On the number of Hamiltonian cycles in a Boolean cube*, Diskretn. Anal. Issled. Oper., **8** (2001) 52–62 (in Russian).

[SVS83] J. Silverman, V.E. Vickers and J.L. Sampson, *Statistical estimates of the n-bit Gray codes by restricted random generation of permutations of 1 to $2^n$*, IEEE Trans. Inform. Theory, **29-6** (1983) 894–901.

[Sl08] N.J.A. Sloane, Ed. (2008), *The On-Line Encyclopedia of Integer Sequences*, published electronically at www.research.att.com/ njas/sequences/, Sequence A066037.

[SQL01] J.Siek, Lie-Quan Lee and A.Lumsdaine, *The Boost Graph Library (BGL)*, `http://www.boost.org/doc/libs/1_42_0/libs/graph/doc/index.html`